\documentclass[reprint,
superscriptaddress,
amsmath,amssymb,aps,showkeys,showpacs,
twoside,final,secnumarabic,%raggedbottom,
nofootinbib]{revtex4-2}

\usepackage[paperwidth=205mm,paperheight=290mm,top=17mm,bottom=25mm,
inner=17mm,outer=17mm,
twoside]{geometry}

\usepackage{cmap} % Улучшенный поиск %русских слов в~полученном pdf-файле
\usepackage[T1,T2A]{fontenc}
\usepackage[utf8]{inputenc}
\usepackage[russian,english]{babel}
\usepackage{color}
\usepackage{graphicx}% Include figure files
\usepackage{dcolumn}% Align table columns on decimal point
\usepackage{bm} % bold math
\usepackage[unicode=true,colorlinks=true,linkcolor=magenta, urlcolor=blue, citecolor = blue,breaklinks]{hyperref}
\usepackage{multirow}
\usepackage{url}
\usepackage{breakurl}
\DeclareGraphicsExtensions{.eps}

\newcount\issue
\newcount\Vol
\newcount\numb
\usepackage{fancyhdr} %this packages %provides fancy up and bottom of page
\def\Vol{\textbf{78}}
\def\numb{x}
\begin{document}

%====== Начало шапки статьи  ============
\title{DARK SECTOR AT BESIII\\[20pt] }
%Insert your title here\\with Forced Linebreak} 

\def\addressa{College of Physics, Jilin University, Changchun 130012, People’s Republic of China}

\author{\firstname{Vindhyawasini}~\surname{Prasad} \\
 (On behalf of the BESIII Collaboration) }
\email[E-mail: ]{vindy@jlu.edu.cn }
\affiliation{\addressa}
%\author{\firstname{A.B.}~\surname{Petrov}}
%\affiliation{\addressa}
% \author{\firstname{A.B.}~\surname{Sidorov}}
%\affiliation{\addressb}

\received{xx.xx.2025}
\revised{xx.xx.2025}
\accepted{xx.xx.2025}

\begin{abstract}
Dark matter (DM) plays a crucial role in explaining the observed astrophysical anomalies, galaxy rotation curves, and other fundamental characteristics of  the Universe. Many extensions of the Standard Model (SM), such as the dark hidden-sector model, provide an attractive framework in which DM can couple to SM particles via various portals. These portals give rise to several possible new particles beyond the SM, such as light Higgs bosons, dark photons, and axion-like particles. Furthermore, the origin of DM and the observed asymmetry between visible matter and antimatter may be connected through the introduction of a dark baryon. If the masses of these hypothetical particles lie in the few-GeV range, they can be explored at high-intensity $e^+e^-$ collider experiments, such as the BESIII experiment. This report presents a review of recent BESIII results related to searches for dark-sector particles.

\end{abstract}

\pacs{Suggested PACS}\par
\keywords{Suggested keywords   \\[5pt]}
%DOI:  

\maketitle
\thispagestyle{fancy}

%====== Начало  статьи  ============

\section{Introduction}\label{intro}
The Standard Model (SM) of particle physics has been remarkably successful~\cite{SM}, and the discovery of the Higgs boson in 2012 completed its particle spectrum~\cite{Higgsdisc}. However, it is not considered a complete theory due to several unresolved issues from both experimental and theoretical sides. Experimentally, anomalies such as  the muon $g-2$~\cite{mug-2} and tensions in $W$-boson mass measurement~\cite{wmass}  have challenged the completeness of the SM. Theoretically, the SM can not explain the observed baryon asymmetry of the universe, neutrino masses and oscillations, or the existence of dark matter (DM)~\cite{DM}. Consequently, extensions of the SM are desirable to address these shortcomings.  

DM accounts for a dominant contribution to the total matter density of the universe and is completely non-interacting with the strong and electromagnetic interactions. Its presence so far has been inferred solely from gravitational effects. Therefore, the nature of DM remains elusive. A precise understanding of DM is desirable to explain the features of recent astrophysical observations~\cite{anomalies} and the total matter density of the universe. Asymmetric DM models proposing GeV-scale DM particles carrying baryon number offer an attractive framework for simultaneously addressing the source of the DM and matter–antimatter asymmetry of the universe~\cite{DB1, DB2}. Additionally, light DM particles could enhance the decay rates of flavor-changing neutral current (FCNC) processes~\cite{JusakTand}, which are highly suppressed by  the Glashow Iliopoulos-Maiani mechanism~\cite{GIM}, making them accessible to the current generation of collider experiments.

Despite overwhelming evidence of DM via astrophysical observations~\cite{anomalies}, the interaction of the DM with the SM particles remains unknown. Many models beyond the SM, such as dark hidden sector model~\cite{Arkani}, introduce “portals” that mediate interactions between DM and SM particles, predicting new particles such as light Higgs bosons, dark photons, axion-like particles (ALPs), and sterile neutrinos~\cite{Essig}. If these particles have masses in the MeV–GeV range, they can be probed in high-intensity electron-positron collider experiments, such as BESIII~\cite{bes3}.  BESIII~\cite{bes3} is a China-based symmetric $e^+e^-$ collider experiment that has accumulated a huge amount of data samples at several energy points from 2.0 to 4.95 GeV, including the $J/\psi$, $\psi(2S)$, and $\psi(3770)$ resonances. These large data samples have enabled BESIII to explore the possibility of various flavors of DM particles, including searches for a light Higgs boson~\cite{lightHiggs}, axion-like particles (ALPs)~\cite{ALP}, massive~\cite{massiveDP} and massless~\cite{masslessDP, ctougamma, Sigma} dark photons, invisible decays of the $K_S^0$ meson~\cite{InvisKs}, dark baryon~\cite{DB}, and muon-philic vector or scalar boson~\cite{muonphilic}. This report reviews recent results from the BESIII experiment related to DM particle searches.

\section{Search for a light Higgs boson}
A light Higgs boson is predicted by many models beyond the SM, such as the Next-to-Minimal Supersymmetric Standard Model (NMSSM)~\cite{Maniatis}, that extend the Higgs sector of the SM to include additional Higgs fields. The Higgs sector of the NMSSM contains three $CP$-even, two $CP$-odd and two charged Higgs bosons. The mass of the lightest $CP$-odd Higgs boson, $A^0$, could be smaller than twice the mass of the charm quark, thus making it accessible via radiative decays of $J/\psi$~\cite{Wilczek}. The couplings of the Higgs field to up- and down-type quark pairs are  proportional to $\cot\beta$ and $\tan\beta$, respectively, where $\tan\beta$ represents the ratio of the vacuum expectation values of the up- and down-type Higgs doublets. The branching fraction of $J/\psi \to \gamma A^0$ is predicted to be in the range $10^{-9}$–$10^{-7}$ depending on $m_{A^0}$, $\tan\beta$, and other NMSSM parameters~\cite{Dermisek}.

\begin{figure*}
\includegraphics[width=1.0\textwidth]{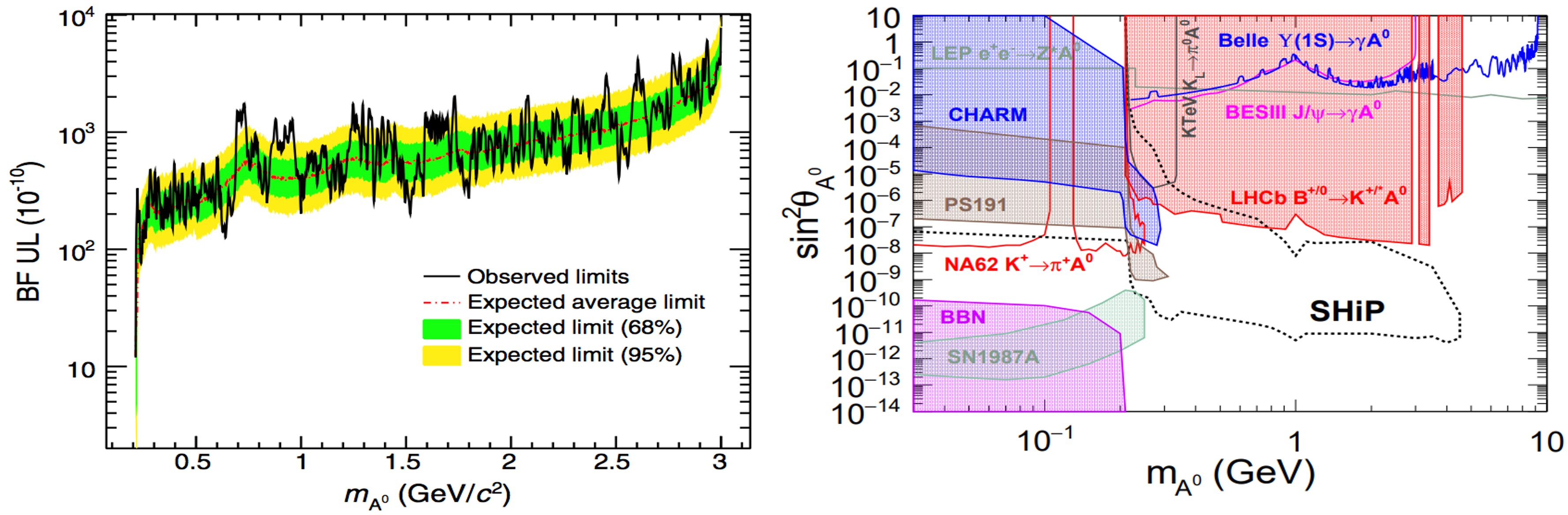}% Here is how to import EPS art
\caption{\label{ULA0} $90\%$ CL upper limits on product branching fractions $\mathcal{B}(J/\psi \to \gamma A^0) \times \mathcal{B}(A^0 \to \mu^+\mu^-)$ (left) and mixing angle ($\sin^2 \theta_{A^0}$) (right) as a function of $A^0$ mass.}
\end{figure*}

Searches for $A^0$ in various final states have been performed by many collider experiments, including BaBar~\cite{BaBar}, Belle~\cite{Belle}, and BESIII~\cite{lightHiggs}. So far only null results have been reported. BESIII  has recently performed a search for dimuon decays of the $A^0$ via radiative $J/\psi$ decays using a dataset of $9.0\times10^{9}$ $J/\psi$ events. No evidence of $A^0$ production is found, and BESIII set $90\%$ confidence-level (CL) upper limits on the product branching fraction $\mathcal{B}(J/\psi \to \gamma A^0)\times\mathcal{B}(A^0 \to \mu^+\mu^-)$ in the range $(1.2-778.0)\times10^{-9}$ for $0.212\le m_{A^0}\le 3.0\ \text{GeV}/c^2$, as shown in Fig.~\ref{ULA0}. The new BESIII result improves upon the previous BESIII measurement by a factor of about 6–7~\cite{lightHiggs}, and it supersheds the BaBar~\cite{BaBar} and Belle~\cite{Belle} measurements in the low-mass region, as shown in the plot of  a mixing angle ($\sin^2 \theta_{A^0}$)  versus $m_{A^0}$ in Fig.~\ref{ULA0}.

\section{Search for axion-like particle}
An axion is a pseudoscalar particle originally introduced through the spontaneous breaking of the Peccei–Quinn symmetry~\cite{Peccei-Quinn} to solve the strong $CP$ problem of QCD~\cite{Strong-CP}. Many models beyond the SM, such as extended Higgs sector~\cite{ExtHiggs} and string theory~\cite{String-th}, predict the existence of axion-like particles (ALPs). The spin-parity of these ALPs is the same as that of the axion, but their masses and coupling characteristics may differ. An ALP couples to a photon pair with a coupling constant $g_{a\gamma\gamma}$. Recently, BESIII has  explored the possibility of di-photon decays of an ALP via radiative $J/\psi \to \gamma a$ decays, using a dataset of 10 billion $J/\psi$ events~\cite{ALP}. This $J/\psi$ data sample is expected to include about $95.6\%$ contribution from radiative $J/\psi \to \gamma a$ decays and about $4.4\%$ contribution from the ALP-strahlung process $e^+e^- \to \gamma a$~\cite{Marlo}, along with dominant SM backgrounds from $J/\psi \to \gamma P$ ($P=\pi^0, \eta, \eta'$)~\cite{PDG}.

\begin{figure*}
\includegraphics[width=1.0\textwidth]{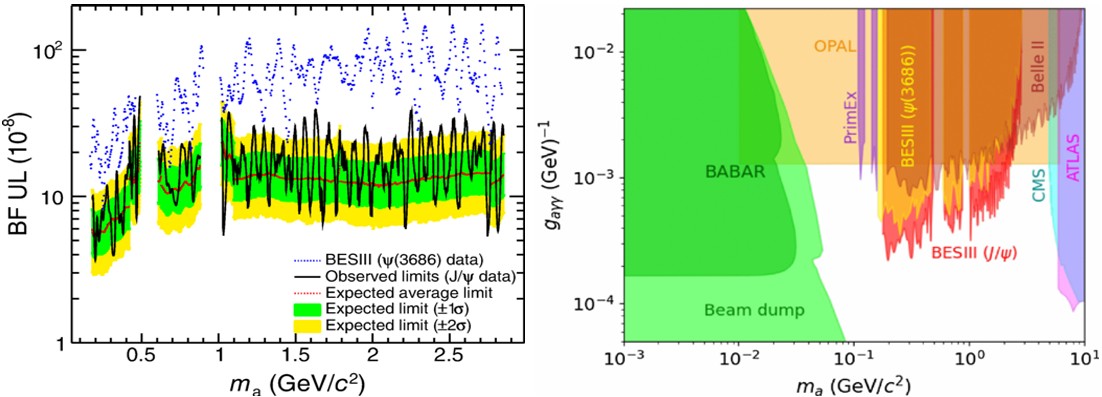}% Here is how to import EPS art
\caption{\label{ALPBF} The $95\%$ CL upper limits on product branching fractions $\mathcal{B}(J/\psi \to \gamma a)\times \mathcal{B}(a \to \gamma\gamma)$ (left) and axion–photon coupling $g_{a\gamma\gamma}$ (right) as a function of $m_a$. These limits are the most stringent to date in the explored mass region.}
\end{figure*}

 The search for ALPs is conducted by performing a series of maximum-likelihood (ML) fits to the di-photon invariant-mass spectrum $m_{\gamma\gamma}$, which includes all  three possible combinations of two-photon pairs after excluding the background contributions from $J/\psi \to \gamma P$. No significant signal  for ALP production is observed, and we set $95\%$ CL upper limits on the product branching fraction $\mathcal{B}(J/\psi \to \gamma a)\times \mathcal{B}(a \to \gamma\gamma)$  and $g_{a \gamma \gamma}$ to be less than  $(3.7-48.5)\times 10^{-8}$ and $(2.2-101.8) \times 10^{4}$, respectively, for the mass range of $0.18 \le m_a \le 2.85$ GeV/$c^2$, as shown in Fig.~\ref{ALPBF}. These exclusion limits are the most stringent to date  in the search mass region~\cite{ALP}.

\section{Search for dark photon}
The dark photon ($\gamma'$) is introduced in the minimum extension of the SM with an additional Abelian group~\cite{Arkani} and serves as a portal between the SM matter and dark sector~\cite{Essig}. The $\gamma'$ can be divided into the following two categories: 1) the massive dark photon, which arises when the symmetry of the additional Abelian gauge group is spontaneously broken and 2) the massless dark photon when the symmetry remains unbroken~\cite{DPSymmetry}. The massive dark photon may couple to the SM matter through a kinetic mixing parameter, defined as $\epsilon = \alpha'/\alpha$, where $\alpha$ and $\alpha'$ are the fine structure constants of the SM and dark sectors, respectively.  Searches for massive dark photon have been conduced by many collider experiments for both its visible and invisible decay modes~\cite{DPreview}. However, only negative results are reported. The massive dark photon searches at BESIII for its visible and invisible decay modes are detailed in Ref.~\cite{massiveDP}. 

On the other hand, the massless dark  photon remains significantly less constrained in comparison to the massive dark photon. The massless dark photon may couple to SM particles via a dimension-six operator, as proposed in Ref.~\cite{dsix}.  The BESIII collaboration has previously explored the possibility of a massless dark photon through the two-body decay of $\Lambda_c^+ \to p + \gamma'$~\cite{masslessDP}. The following two decay modes have also been explored recently by the BESIII experiment to search for massless photon: 

\subsection{Search for massless dark photon via \boldmath{$D^0 \to \omega \gamma'$} and \boldmath{$D^0 \to \gamma \gamma'$}}
The search for a massless dark photon in the FCNC decays $D^0 \to \omega \gamma'$ and $D^0 \to \gamma \gamma'$ has been performed using 7.9 fb$^{-1}$ of $\psi(3770) \to D^0 \bar{D}^0$ data~\cite{ctougamma} , employing the double-tag (DT) technique pioneered by the MARK III Collaboration~\cite{DT}. The decays $D^0 \to \omega \gamma'$ and $D^0 \to \gamma \gamma'$ can proceed through the $cu \gamma'$ effective coupling in dimension-six operators~\cite{dsix}.  In the DT technique, one of the $\bar{D}^0$ mesons is tagged using its dominant hadronic decay modes detailed in Ref.~\cite{DT}, and then the other $D$ meson is reconstructed in the signal mode of interest~\cite{DT}. The branching fractions of these decays are related to $|C|^2 + |C_5|^2$~\cite{ctougamma}, where $C=\Lambda_{\rm NP}^{-2}(C_{12}^U + C_{12}^{U*})\nu/\sqrt{8}$ and $C_5=\Lambda_{\rm NP}^{-2}(C_{12}^U - C_{12}^{U*})\nu/\sqrt{8}$, with the Higgs vacuum expectation value $\nu = 246.2$~GeV. These coefficients are determined by the new physics energy scale $\Lambda_{\rm NP}$ and the up-type dimensionless coefficient $C_{12}^U$.

\begin{figure*}[htbp]
\centering
\includegraphics[width=1.0\textwidth]{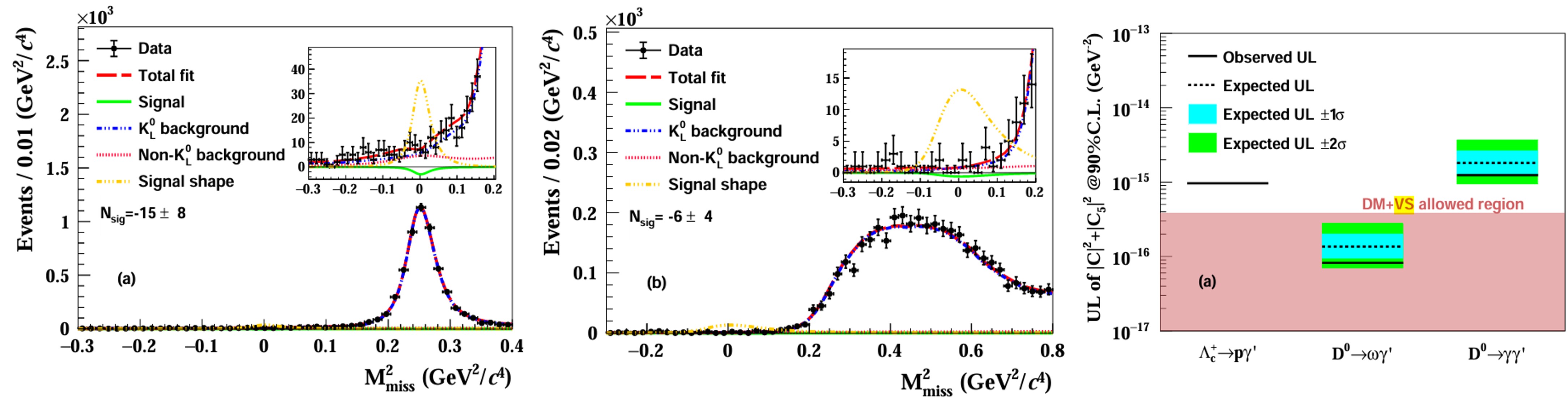}
\caption{\label{MasslessDM} Fits to the missing mass squared $M_{\rm miss}^2$ distributions of the accepted candidates for $D^0 \to \omega \gamma'$ (left) and $D^0 \to \gamma \gamma'$ (middle). The right-hand plot shows the 90\% CL upper limits on $|C|^2 + |C_5|^2$ obtained in this work, compared with the result from $\Lambda_c^+ \to p \gamma'$ in Ref.~\cite{masslessDP}.}
\end{figure*}

The signal events for the massless dark photon are extracted by performing an extended ML fit to the missing mass squared, defined as

\begin{equation}
M_{\rm miss}^2 = |p_{\rm cms} - p_{\bar{D}^0} - p_{\omega(\gamma)}|^2 / c^4,
\end{equation}

\noindent where $p_{\rm cms}$ is the four-momentum of the $e^+e^-$ center-of-mass system in the laboratory frame, $p_{\omega(\gamma)}$ is the kinematically fitted (reconstructed) four-momentum of $\omega$ ($\gamma$), and $p_{\bar{D}^0}$ is the four-momentum of the $\bar{D}^0$ meson. The fit yields signal events of $-15 \pm 8$ for $D^0 \to \omega \gamma'$ and $-6 \pm 4$ for $D^0 \to \gamma \gamma'$, as shown in Fig.~\ref{MasslessDM} (left and middle), both consistent with zero. The upper limits on the branching fractions at the 90\% CL are calculated to be $\mathcal{B}(D^0 \to \omega \gamma') < 1.1 \times 10^{-5}$ and $\mathcal{B}(D^0 \to \gamma \gamma') < 2.0 \times 10^{-6}$ after including the systematic uncertainties. 

Finally, these upper limits are translated into a constraint on the parameter related to the new physics energy scale, $|C|^2 + |C_5|^2$, as shown in Fig.~\ref{MasslessDM} (right). This value is found to be less than $8.2 \times 10^{-17}$~GeV$^{-2}$, reaching for the first time the region allowed by DM and vacuum stability (VS) in the universe, and surpassing the previous limit from the $\Lambda_c^+ \to p \gamma'$ measurement~\cite{masslessDP} by more than an order of magnitude~\cite{ctougamma}. 

\subsection{Search for a Massless Particle in the \boldmath{$\Sigma^+ \to p + {\rm invisible}$} Decay}

The FCNC decay $\Sigma^+ \to p + {\rm invisible}$ can proceed from the loop-induced quark transition $s \to d \nu \bar{\nu}$, which is strongly suppressed by the  GIM mechanism~\cite{GIM}. The branching fraction of such decays is predicted by the SM to be less than $10^{-11}$~\cite{Jusak}. However, in the presence of new physics beyond the SM, the branching fractions of certain FCNC hyperon decays could be enhanced up to the level of $10^{-4}$~\cite{Su}. The possible new invisible particle could be either a QCD axion or a massless dark photon. The QCD axion is expected to have a mass smaller than 1~eV and a lifetime longer than the age of the universe. Therefore, the search for a massless particle through hyperon decays provides a sensitive probe for new physics beyond the SM.

The search for a massless particle in $\Sigma^+ \to p + {\rm invisible}$ has been performed using about $10^7$ $\Sigma^+ \bar{\Sigma}^-$ events selected from $(10087 \pm 0.44) \times 10^{10}$ $J/\psi$ events collected by the BESIII detector, employing a DT technique~\cite{Sigma}. One $\bar{\Sigma}^-$ candidate from the decay $J/\psi \to \Sigma^+ \bar{\Sigma}^-$ is reconstructed via its dominant decay mode $\bar{\Sigma}^- \to \bar{p} \pi^0$. The other $\Sigma^+$ is allowed to decay via $\Sigma^+ \to p + {\rm invisible}$, where the mass of the invisible particle is constrained to be zero by performing a kinematic fit.

The signal yield is extracted by fitting the distribution of the total energy deposited in the electromagnetic calorimeter (EMC) by extra photons ($E_{\rm extra}$) that are not used in $\pi^0 \to \gamma\gamma$ reconstruction. The $E_{\rm extra}$ distribution peaks near zero for signal-like events, while it deviates from zero for background processes. The main background contributions in $E_{\rm extra}$ distribution arise from events containing extra $\pi^0$ or $\gamma$ candidates ($E_{\rm extra}^{\rm DT\,\pi^0/\gamma}$) and other sources ($E_{\rm extra}^{\rm other}$), including detector noise and unrelated events. The shape of $E_{\rm extra}^{\rm DT\,\pi^0/\gamma}$ is determined from Monte Carlo simulation, while the shape of $E_{\rm extra}^{\rm other}$ is obtained using a data-driven method.

\begin{figure*}[htbp]
\centering
\includegraphics[width=1.0\textwidth]{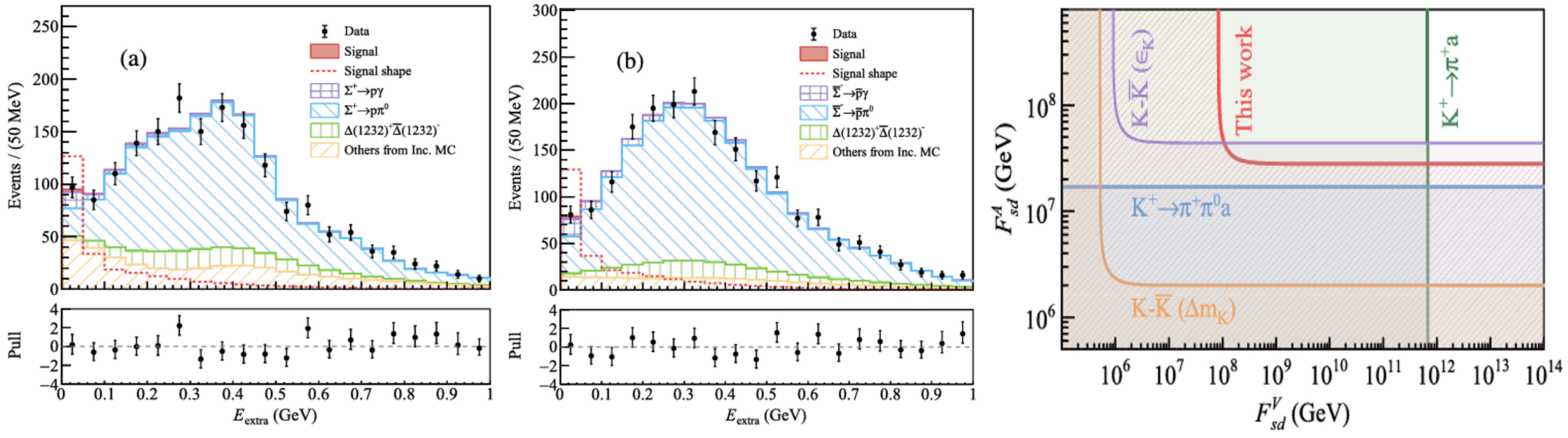}
\caption{\label{sigmapnunu} Fits to the $E_{\rm extra}$ distributions for (left) $\Sigma^+ \to p + {\rm invisible}$ and (middle) $\bar{\Sigma}^- \to \bar{p} + {\rm invisible}$ DT signal channels. (Right) The 90\% C.L. exclusion limit on the $s \to d$ axion–fermion effective coupling obtained from this analysis (hatched region excluded), compared with existing constraints from $K^+ \to \pi^+ a$, $K^+ \to \pi^+ \pi^0 a$, $K$–$\bar{K}$ mixing ($\Delta m_K$), and the CP-violating parameter $\epsilon_K$.}
\end{figure*}

No evidence for $\Sigma^+ \to p + {\rm invisible}$ is observed, as shown in Fig.~\ref{sigmapnunu} (left and middle). The upper limit on the branching fraction $\mathcal{B}(\Sigma^+ \to p + {\rm invisible})$ is determined to be $3.2 \times 10^{-5}$ at the 90\% CL. This limit has been translated into the axial–vector component ($F_{sd}^A$) of axion-fermion coupling, as shown in Fig.~\ref{sigmapnunu} (right). This limit shows significantly improvement over the bound from $K$–$\bar{K}$ mixing ($\Delta m_K$) and is comparable to the limits from searches for $K^+ \to \pi^+ \pi^0 a$ and measurements of the CP-violating parameter $\epsilon_K$ in the kaon system~\cite{Sigma}.

\section{Invisible Decays of the \boldmath{$K_S^0$} Meson}

In the SM, the FCNC decay $K_S^0 \to \nu \bar{\nu}$ proceeds via the $s \to d \nu \bar{\nu}$ transition. This decay is highly suppressed due to angular momentum conservation and remains suppressed even in the case of massive neutrinos because of the unfavorable helicity configuration, resulting in a predicted branching fraction of less than $10^{-16}$~\cite{Kstoinvis}. Consequently, the search for invisible decays of the $K_S^0$ meson provides a sensitive test of the SM. Contributions from new physics models, such as the Two-Higgs-Doublet Model (2HDM), predict that the branching fraction for invisible decays of the $K_S^0$ meson could reach the level of $10^{-4}$~\cite{Kstoinvis}. Additionally, theories such as the mirror-matter model, which postulate the existence of a mirror world parallel to our own, suggest that the $K_S^0$ invisible decay could occur at the order of $10^{-6}$~\cite{mirror}. Furthermore, invisible decays of $K_S^0$ can provide valuable input for tests of charge, parity, and time-reversal (CPT) symmetry, since the Bell–Steinberger relation connects CPT violation to the amplitudes of all decay channels of neutral kaons, under the assumption that no invisible modes currently exist~\cite{InvisKs}.

\begin{figure*}[htbp]
\centering
\includegraphics[width=1.0\textwidth]{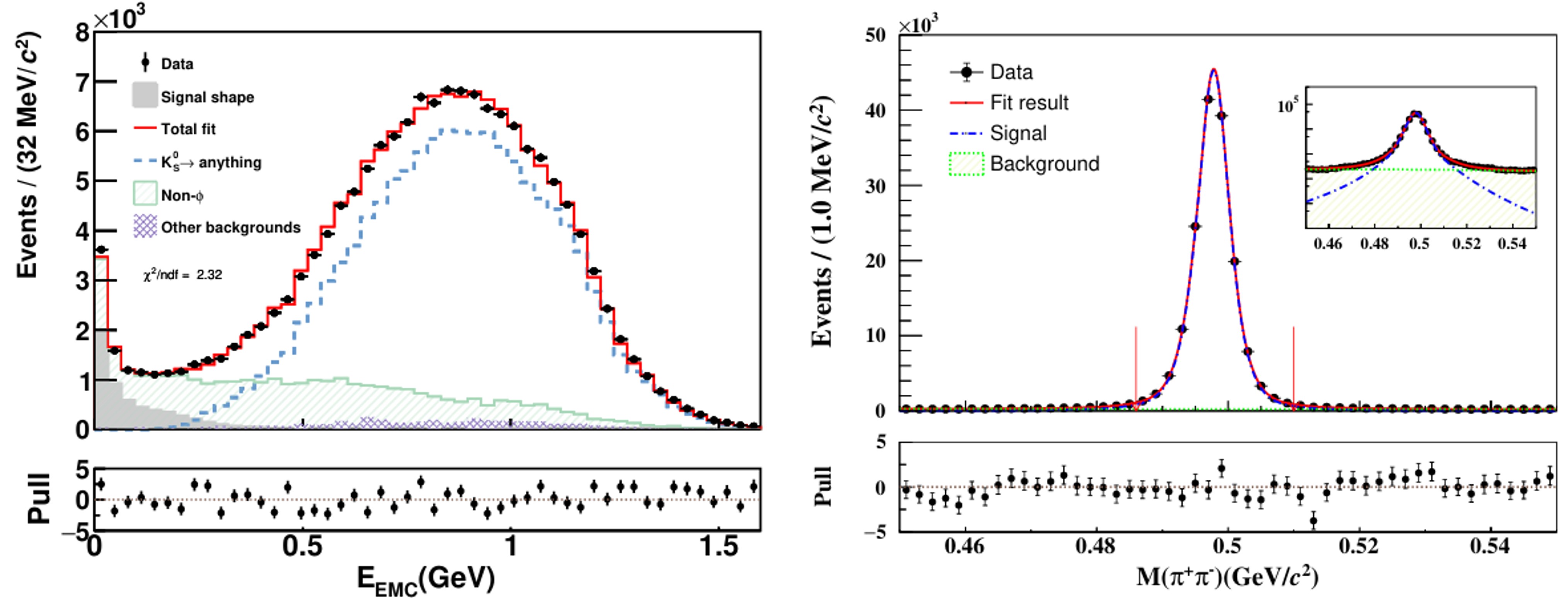}
\caption{\label{fig:KSinvisible} Fits to the $E_{\rm EMC}$ distributions used to extract signal events for $K_S^0 \to {\rm invisible}$ (left) and the di-pion invariant mass spectrum from visible $\pi^+\pi^-$ decay  from the normalization channel (right). Backgrounds from four-pion and non-$\phi$ processes are subtracted from the signal yield shown in the right-hand figure.}
\end{figure*}

The first direct search for invisible decays of the $K_S^0$ meson has been performed by the BESIII Collaboration using $10^{10}$ $J/\psi$ events~\cite{InvisKs}. The $K_S^0$ candidates are selected from the decay $J/\psi \to \phi K_S^0 K_S^0$, which exhibits a relatively low background contribution because the decay $J/\psi \to \phi K_S^0 K_L^0$ is forbidden by C-parity conservation. In this search, the $\phi$ meson is reconstructed via $\phi \to K^+K^-$, and one of the $K_S^0$ mesons is reconstructed through $K_S^0 \to \pi^+\pi^-$, while the other $K_S^0$ meson is used to search for the invisible decay.  The distribution of the energy deposited in the electromagnetic calorimeter ($E_{\rm EMC}$) is used to distinguish signal from background. Since invisible particles do not interact with detector material, the $E_{\rm EMC}$ distribution is expected to peak near zero, whereas background events deviate from zero, as seen in Fig.~\ref{fig:KSinvisible} (left). To reduce systematic uncertainties, the $J/\psi \to \phi K_S^0 K_S^0$ decay is used as the normalization,  where both $K_S^0$ candidates are allowed to reconstruct from $K_S^0 \to \pi^+\pi^-$, after subtracting four-pion and non-$\phi$ background contributions. Figure~\ref{fig:KSinvisible} (right) shows the  di-pion invariant mass spectrum from visible $K_S^0 \to \pi^+\pi^-$ decay  from the normalization channel. No evidence of  significant signal events is observed, and the upper limit on the branching fraction of the invisible decays of the $K_S^0$ meson is determined to be less than $8.4 \times 10^{-4}$ at the 90\% CL for the first time. 

\section{Search for dark baryon in \boldmath{$\Xi^- \to \pi^- + {\rm invisible}$} decay}

The similarity between the dark matter and baryon energy densities, $\rho_{\rm DM} \approx \rho_{\rm baryon}$~\cite{rhoDM}, suggests a possible connection between their origins, implying that dark matter may carry a non-zero baryon number~\cite{DB1, DB2}. The baryonic dark sector is further motivated by the long-standing discrepancy between neutron lifetimes measured in beam and bottle experiments, which could be resolved if the neutron decays into DM states carrying baryon number with a branching fraction at the level of $1\%$~\cite{Fornal}. The $B$-mesogenesis mechanism can explain both the asymmetry between visible matter and antimatter and the origin of dark matter~\cite{mesogenesis}. The existence of dark-sector antibaryons has been explored in $B$-meson decays at the BaBar experiment~\cite{DBBabar} and, more recently, in invisible decays of the $\Lambda$ baryon by the BESIII experiment~\cite{invLambda}. Complementary to these searches, hyperons offer an opportunity to probe the baryonic dark sector via decays into final states containing dark baryons, which would appear as missing energy in the detector.

A search for a dark baryon $\chi$ in the two-body decay $\Xi^- \to \pi^- + {\rm invisible}$ is performed using approximately $10^7$ events produced from $J/\psi \to \Xi^- \bar{\Xi}^+$ decays, based on $(10.0 \times 10^9)$ $J/\psi$ events collected with the BESIII detector~\cite{DB}. In the process $J/\psi \to \Xi^- \bar{\Xi}^+$, the $\Xi^-$ candidate is reconstructed by tagging a $\bar{\Xi}^+$ that decays via $\bar{\Xi}^+ \to \pi^+ \bar{\Lambda}(\to \bar{p}\pi^+)$ on the recoil side. The dark baryon $\chi$ is inferred as an invisible particle, with the search performed at assumed masses of 1.07, 1.10, $m_{\Lambda}$, 1.13, and 1.16~GeV/$c^2$, where $m_{\Lambda}$ is the $\Lambda$-baryon mass. 

\begin{figure*}[htbp]
\centering
\includegraphics[width=1.0\textwidth]{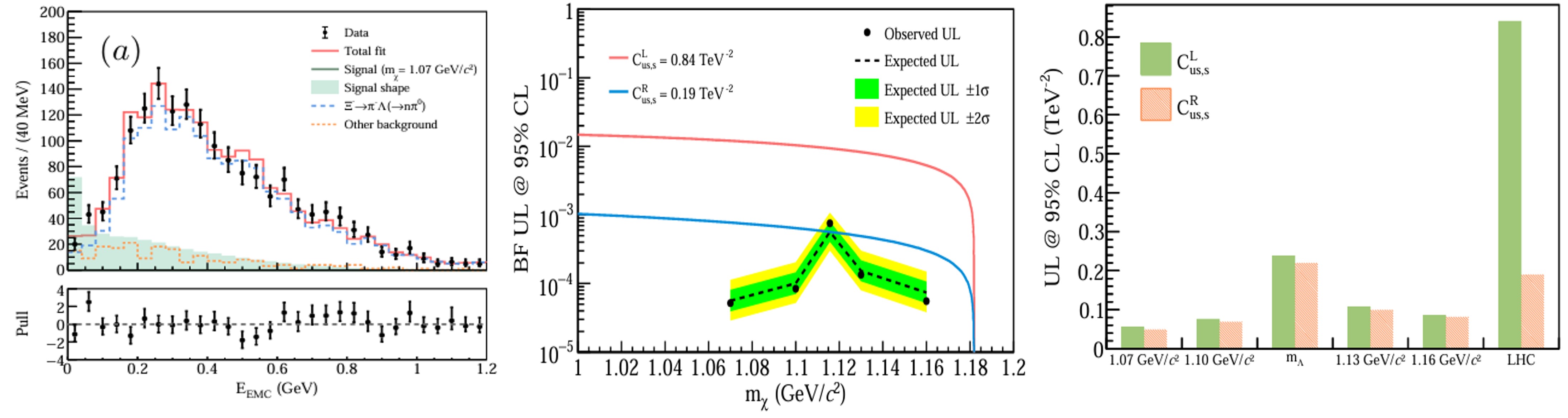}
\caption{\label{fig:Chimass} Fits to the $E_{\rm EMC}$ distributions used to extract signal events for $\Xi^- \to \pi^- + {\rm invisible}$ (left), observed (black points) and expected (yellow and green bands) 95\% CL upper limits on $\mathcal{B}(\Xi^- \to \pi^- + {\rm invisible})$ under different $m_{\chi}$ hypotheses (middle), and 95\% CL upper limits on the Wilson coefficients $C_{us,s}^L$ and $C_{us,s}^R$ under different $m_{\chi}$ assumptions, including constraints derived from LHC searches (right).}
\end{figure*}

The invisible signal is expected to produce an $E_{\rm EMC}$ distribution peaking near zero for signal-like events and deviating from zero for background-like events such as $\Xi^- \to \pi^- \Lambda(\to n \pi^0)$, as shown in Fig.~\ref{fig:Chimass}~(left). No significant signal is observed, and 90\% (95\%) CL upper limits on the branching fraction $\mathcal{B}(\Xi^- \to \pi^- + {\rm invisible})$ are set as a function of the $\chi$ mass, as shown in Fig.~\ref{fig:Chimass}~(middle). The branching fraction of the $\Xi^- \to \pi^- \chi$ decay is related to the left-handed and right-handed effective operators, parameterized by the Wilson coefficients $C_{us,s}^L$ and $C_{us,s}^R$, respectively. The 95\% CL upper limits on the corresponding Wilson coefficients are determined to be $C_{us,s}^L < 5.5 \times 10^{-2}$~TeV$^{-2}$ and $C_{us,s}^R < 4.9 \times 10^{-2}$~TeV$^{-2}$. The obtained constraints on $C_{us,s}^L$ and $C_{us,s}^R$ are more stringent than the previous limits from LHC searches for colored mediators~\cite{DB}, as shown in Fig.~\ref{fig:Chimass}~(right).

\section{Search for invisible muon-philic scalar or vector boson via \boldmath{$J/\psi \to \mu^+\mu^- + {\rm invisible}$}}

A new type of massive vector boson ($X_1$) or scalar boson ($X_0$) may appear in SM extensions such as the anomaly-free gauged $U(1)$ or $U(1)_{L_{\mu}-L_{\tau}}$ models. These particles couple only to the second and third generations of leptons ($\mu, \nu_{\mu}, \tau, \nu_{\tau}$) with coupling strengths $g'_{0,1}$~\cite{muphilic}. The $X_{0,1}$ can contribute to the muon anomalous magnetic moment and potentially explain the muon $(g-2)$ anomaly~\cite{mug-2} . They can be produced via $J/\psi \to \mu^+\mu^- X_{0,1}$. The current experimental bounds on such massive vector or scalar bosons mainly come from the BaBar and CMS experiments for muon-pair final states, and from the Belle~II experiment for invisible final states~\cite{muExpt}.

Searches for a light muon-philic scalar $X_0$ or vector $X_1$ have been performed via $J/\psi \to \mu^+\mu^- X_{0,1}$ with $X_{0,1}$ decaying invisibly, using $(8.998 \pm 0.039) \times 10^9$ $J/\psi$ events collected by the BESIII detector~\cite{muonphilic}. The signal yields for $M(X_{0,1}) < 0.4$~GeV/$c^2$ are extracted by performing a series of extended maximum-likelihood fits to the recoil mass of the dimuon system, defined as
\begin{equation}
M_{\rm recoil}^2(\mu^+\mu^-) = (p_{J/\psi} - p_{\mu^+} - p_{\mu^-})^2,
\end{equation}
where $p_{J/\psi}$ and $p_{\mu^{\pm}}$ are the four-momenta of the $J/\psi$ and $\mu^{\pm}$ particles, respectively. The dominant backgrounds in this mass region mainly come from $J/\psi \to \mu^+\mu^-(\gamma)$ and $e^+e^- \to \mu^+\mu^-(\gamma)$ processes, where the photons arise from initial or final state radiation. 

The signal yields for the higher mass region, $0.4 < M(X_{0,1}) < 1.0$~GeV/$c^2$, are obtained by fitting the $M_{\rm recoil}$ distribution. No significant signal is observed, and the upper limits on the coupling constants $g'_{0,1}$ between the muon and the $X_{0,1}$ particles are set in the range $1.1 \times 10^{-3}$ to $1.0 \times 10^{-2}$ for $1 < M(X_{0,1}) < 1000$~MeV/$c^2$ at 90\%  CL. The obtained limits are more stringent than the Belle~II results in the mass range $200$--$860$~MeV/$c^2$, as detailed in Ref.~\cite{muonphilic}.

\section{Summary}
Based on data samples collected at several energy points between 2 and 4.9~GeV, the BESIII experiment has performed the searches  for various dark-sector particles predicted by many models beyond the SM. These searches include  light Higgs boson~\cite{lightHiggs}, axion-like particles (ALPs)~\cite{ALP}, massive~\cite{massiveDP} and massless~\cite{masslessDP, ctougamma, Sigma} dark photons, invisible decays of the $K_S^0$ meson~\cite{InvisKs}, dark baryon~\cite{DB}, and muon-philic vector or scalar boson~\cite{muonphilic}. Although these searches have reported null results so far, their obtained exclusion limits can constrain a large fraction of the parameter space of the new physics models beyond the SM. More results are expected to be released in the near future, especially with the recently collected 20~fb$^{-1}$ of BESIII $\psi(3770)$ data.

\begin{acknowledgments}
We wish to acknowledge the organizers of the LOMONOSOV 2025 conference for their wonderful hospitality and for providing a stimulating environment for physics discussions.

\end{acknowledgments}

\section*{FUNDING}
This work was supported by the Seed Funding of Jilin University.

\section*{CONFLICT OF INTEREST}
The authors declare that they have no conflicts of interest.

% The \nocite command causes all entries in a bibliography to be printed out
% whether or not they are actually referenced in the text. This is appropriate
% for the sample file to show the different styles of references, but authors
% most likely will not want to use it.
\nocite{*}

%%%%%%%%%%%%%%%%%%%%%%%%%%%%%%%%
% USE thebibliography
%%%%%%%%%%%%%%%%%%%%%%%%%%%%%%%%

\end{document}